\documentclass[a4paper]{jpconf}
\usepackage{graphicx,amsmath,amssymb}
\usepackage{iwsmihead}
\pdfoutput=1
\begin{document}

\title{The solution space of metabolic networks: producibility, robustness and fluctuations}

\author{A. De Martino$^{1,2}$, E. Marinari$^2$}

\address{$^1$CNR/INFM and $^2$Dipartimento di Fisica, Sapienza Universit\`a di Roma, p.le A. Moro 2, 00185 Roma (Italy)}

\ead{andrea.demartino@roma1.infn.it,enzo.marinari@roma1.infn.it}

\begin{abstract}
Flux analysis is a class of constraint-based approaches to the study
of biochemical reaction networks: they are based on determining the
reaction flux configurations compatible with given stoichiometric and
thermodynamic constraints.  One of its main areas of application is
the study of cellular metabolic networks. We briefly and selectively
review the main approaches to this problem and then, building on
recent work, we provide a characterization of the productive
capabilities of the metabolic network of the bacterium {\it E.coli} in
a specified growth medium in terms of the producible biochemical
species. While a robust and physiologically meaningful production
profile clearly emerges (including biomass components, biomass
products, waste etc.), the underlying constraints still allow for
significant fluctuations even in key metabolites like ATP and, as a
consequence, apparently lay the ground for very different growth
scenarios.
\end{abstract}

\section{Introduction}

At a rough conceptual level, living cells can be seen as devices that
convey the free energy derived from the breakdown of nutrients (mostly
sugars) into the chemical energy that fuels the production of all the
molecules required for survival and, when possible, growth and
reproduction. The complex intracellular machinery that underlies the
energy transduction is being increasingly unveiled at both the
biochemical (reactions) and the regulatory level (enzymes, their
corresponding genes, the genes' transcription factors, etc) through
the massive genomic information available for different organisms
\cite{gold}. In particular, the network of biochemical processes
collectively known as `metabolism' by which the nutrients are degraded
and the housekeeping molecules are manufactured has been mapped in
great detail for many bacteria and several eukaryotes
\cite{pals}. While the study of the structure and dynamics of single
metabolic pathways has a long standing history in biophysical
chemistry \cite{hs,beard}, the current wealth of data allows to
analyze the behavior of cellular reaction networks at the scale of the
whole genome. This is a crucial step both to shed light on the
emergence of a metabolic phenotype from the underlying genotype and to
formulate testable \cite{sau} predictions on a cell's adaptation and
response to perturbations (the key to multiple biotechnological
applications).

Ultimately, metabolic activity and capabilities are determined (or
limited) by various constraints of chemical (stoichiometric),
thermodynamic and regulatory origin. The uncertainty about their
details increases considerably as one passes from the stoichiometric
to the regulatory level. Current theoretical approaches therefore try
to infer the global organization of metabolism from simple schemes
that implement explicitly only the best known restrictions, making as
few assumptions as possible on the rest. Even so, it turns out that
some predictive and explanatory power on a cell's biochemical
functioning can be achieved.

We begin by presenting a short selective review of stoichiometric
models of metabolic networks as constraint-based systems. We then
focus on one specific problem, that of metabolite producibility,
characterizing the global metabolic output for the bacterium {\it
  E.coli} in a specified growth medium. While a robust and
biologically significant production profile emerges, fluctuations in
nutrient usage or in the production level of key metabolites
occur. Such fluctuations can be associated to different patterns of
metabolic pathway activation. Finally we discuss some of the issues
where statistical mechanics tools may have a significant impact in the
near future. We shall limit biochemical details to a minimum and treat
the reaction system as a standard input-output network. The interested
reader is referred to e.g. \cite{hs} for a thorough introduction to
cellular metabolism.

\section{Flux analysis: a bird's eye view}

\begin{figure}
\begin{center}
\includegraphics[width=15cm]{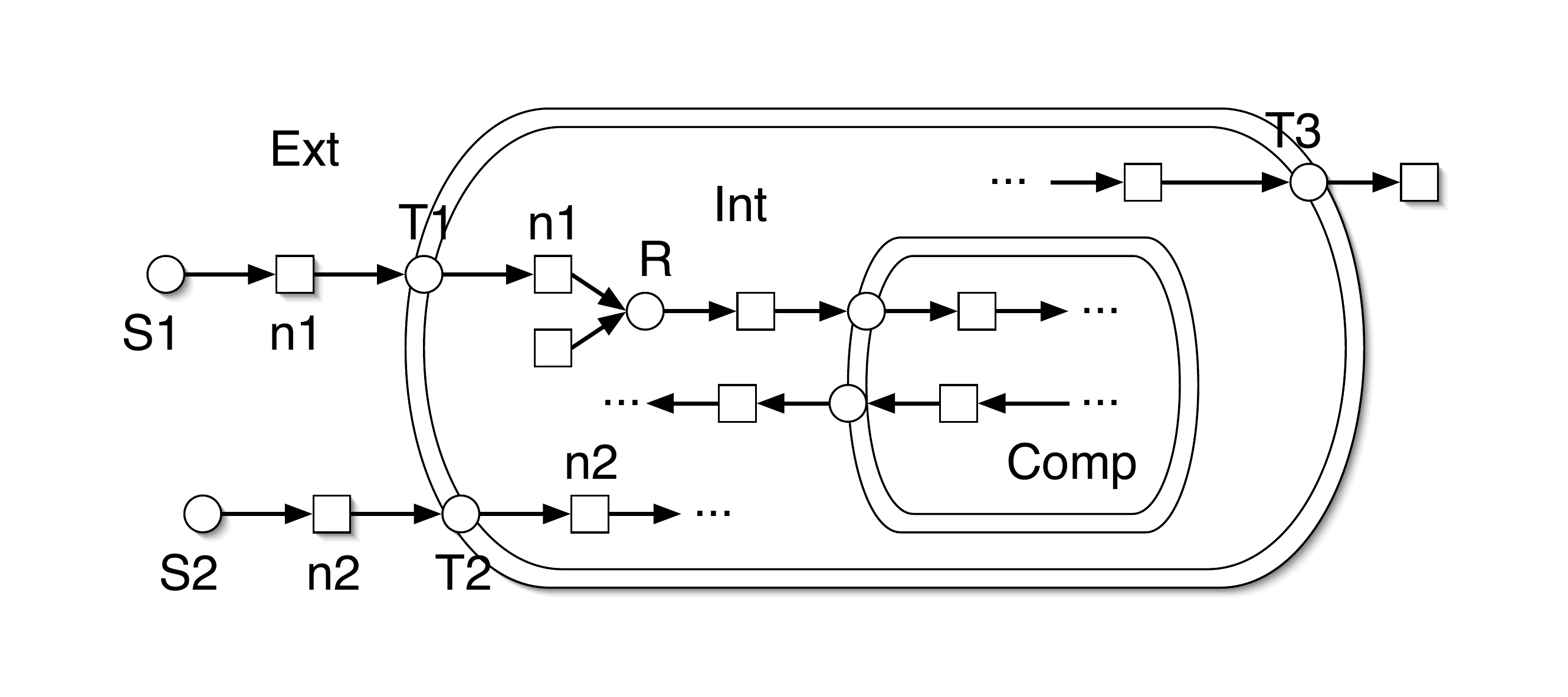}
\caption{\label{cell}Basic scheme of a cellular metabolic network. Ext
  and Int denote respectively the exterior and the interior of the
  cell. Reactions (resp. metabolites) are denoted by circles
  (resp. squares). S1 and S2 are auxiliary fluxes supplying nutrients
  n1 and n2 to the environment. T1, T2 and T3 denote membrane
  transport reactions by which metabolites are taken in or expelled
  from the cell. R is an intracellular reaction. A cellular
  compartment Comp is also shown, together with the corresponding
  transport reactions connecting its interior with the cytosol.}
\end{center}
\end{figure}

In essence (see Fig. \ref{cell}), a cell's metabolic network can be
seen as a set of interconnected chemical reactions coupled with a set
of transport processes.
For bacteria, this includes the membrane transport mechanisms by which
nutrients are brought into the cell, and the intracellular reactions
by which they are degraded and new biochemical species are
produced. In organisms with compartmentalized cells (e.g. eukaryotes)
one should also account for the transport of metabolites into and out
of each compartment, i.e. for the cell's actual geometric
structure. The basic and most reliable information on the, say, $N$
reactions involving $M$ chemical species is encoded in the $M\times N$
{\it stoichiometric matrix} $\mathbf{\Xi}$, whose entry $\xi_i^\mu$
represents the stoichiometric coefficient with which species $\mu$
participates in reaction $i$. $\mathbf{\Xi}$ is a sparse, integer
matrix and a sign convention is usually adopted to discern products
($\xi_i^\mu>0$) from substrates ($\xi_i^\mu<0$). The stoichiometric
matrix often also contains a motivated assumption on reaction
directionality to account for the fact that, while all reactions are
in principle reversible, under physiological conditions some of them
may occur in one direction only. We shall treat physiologically
reversible reactions as two separate processes. In addition, we shall
always assume that $\mathbf{\Xi}$ also includes external supply fluxes
for the nutrients, i.e. auxiliary (free or fixed) reactions that
provide the nutrient to the environment. To have an idea, in such a
setting for the bacterium {\it E.coli} $N\simeq 1100$ and $M\simeq
700$, whereas $N\simeq 1500$ and $M\simeq 900$ for the unicellular
eukaryote {\it S.cerevisi\ae}.

If we denote by $\boldsymbol{\nu}\geq \boldsymbol{0}$ the
$N$-dimensional vector of reaction fluxes (the sign constraint arising
from our treatment of reversibility), the time evolution of the vector
$\mathbf{c}$ of metabolite concentrations is described simply by
$\dot{\mathbf{c}}=\mathbf{\Xi}\boldsymbol{\nu}$. Fluxes depend in
principle on the membrane transport mechanisms or on the enzyme
kinetics through various parameters like the rate constants
$\mathbf{k}$ and $\mathbf{c}$ itself and in turn, through them, on
temperature, activation energy, etc. If such details are known, then
functional dependencies
$\boldsymbol{\nu}=\boldsymbol{\nu}(\mathbf{c},\mathbf{k},\ldots)$ can
be chosen according to the corresponding kinetics
(e.g. Michaelis-Menten or Hill) and one can in principle solve the
dynamical system for the concentrations, provided all underlying
constants are known. Unfortunately, this is rarely the case in
genome-scale networks (see \cite{dyn} for a positive example). While
methods have been developed to partially overcome this limitation and
provide dynamical characterizations of metabolic activity (see
e.g. \cite{dfba,skm}), the standard modeling route assumes, on the
basis of the timescale separation between chemical processes and
genetic regulation, that metabolic networks operate in a non
equilibrium steady state where $\dot{\mathbf{c}}=\mathbf{0}$. Note
that flux vectors satisfying $\mathbf{\Xi}\boldsymbol{\nu}=\mathbf{0}$
describe states in which intracellular metabolites obey Kirchhoff-type
mass-balance conditions. If $N>M$, as it usually occurs in real
metabolic networks, such vectors form a set of dimension
$N-\text{rank}(\mathbf{\Xi})$ embedded in $\mathbb{R}_0^+$ that, in
absence of additional constraints, contains equivalent feasible flux
states of the network. In principle, this set should be explored
uniformly to extract the relevant biological
information. Unfortunately, Monte Carlo sampling becomes unaffordable
as soon as the dimension of the solution space exceeds a few tens (see
\cite{mc} for a feasible case; message-passing algorithms have been
recently proposed as a working alternative for microbial cells
\cite{mul}). In many cases it is however possible to select relevant
configurations by imposing the maximization of an objective function
normally represented as a linear combination of the fluxes with given
coefficients $\boldsymbol{\alpha}$,
i.e. $\boldsymbol{\alpha}\cdot\boldsymbol{\nu}$. This reduces the
problem to a linear optimization one, namely
\begin{equation}
\max_{\boldsymbol{0}\leq\boldsymbol{\nu}\leq\boldsymbol{\nu}_{max}} (\boldsymbol{\alpha}\cdot\boldsymbol{\nu})~~~\text{subject to}~~~\mathbf{\Xi}\boldsymbol{\nu}=\mathbf{0}
\end{equation}
The upper bounds on fluxes can be chosen to either model specific
extracellular conditions (e.g. a certain nutrient's supply or intake
cannot exceed a fixed level), to simulate known physiological
limitations (e.g. the rate of a certain reaction cannot exceed a fixed
level because of limited enzyme availability), or simply to impose
well-defined bounds on the solution space. This scenario is the basis
of various stoichiometry-based approaches such as Flux-Balance
Analysis (FBA) \cite{kau}. Clearly, the vector $\boldsymbol{\alpha}$
contains the crucial biological assumptions. In some cases, it is
grounded in experimental evidence. For instance, bacteria in rich
environments appear to evolve, under selective pressure, to optimize
their growth rate and reproduce as fast as possible \cite{ib}. A good
proxy for a bacterium's growth capability lies in its ability to
generate {\it biomass}, a combination of different metabolites in
precise stoichiometric proportions (including\footnote{See
  \cite{biom2} for a list of metabolite abbreviations.} the 20
proteinogenic amino acids, the molecular energy carrier ATP, water,
key cofactors like nad, nadp and coenzyme-A etc.) that is used to
produce ADP, inorganic phosphate (PI), pyrophosphate (PPI) and water
\cite{biom,biom2}. This reaction represents the cell's use of
metabolic products in macromolecular processes (like building
proteins, membranes, etc.) that are not accounted for in
$\mathbf{\Xi}$. The maximization of biomass yield is in these cases a
widely used criterion implemented by adding the auxiliary biomass
reaction to $\mathbf{\Xi}$. In other cases, the choice may have
physiological justifications.  The need for energetically-efficient
housekeeping e.g. in nutrient-limited conditions suggests that cells
optimize the ATP yield, which requires to maximize the total flux of
all ATP-producing reactions \cite{rama}. Other examples include the
minimization of glucose consumption (a proxy for efficient nutrient
usage \cite{oli}) or of the total flux of intracellular reactions (for
maximal enzymatic efficiency \cite{bona}). We refer the reader to
\cite{sys} for a comparison of the performance of several different
objective functions for predicting the fluxes of the main carbon
pathways in {\it E.coli} and to \cite{sri} for a study of objective
function selection criteria.

This kind of approach can be extended to biologically more complicated
situations like those induced by gene knock-outs that prevent the
execution of certain reactions. Based on experimental evidences, such
scenarios lead to consider different selection criteria, like the
minimization of the overall flux rearrangements with respect to the
wild type (by quadratic programming \cite{segre}) or the minimization
of rearrangements in the large-flux backbone with respect to the wild
type (by mixed linear-integer programming \cite{room}). The former
appears to be suited to capture the transient sub-optimal growth
states that a bacterium takes on immediately after the perturbation,
whereas the latter provides a better description of the states of fast
growth that the knock-out organism reaches on longer time scales under
selective pressure. However the dynamical structure of the response of
metabolic networks to perturbations is far from being satisfactorily
understood.

\section{Producibility and Von Neumann's problem}

An important problem related to flux analysis concerns the link
between the network's structure and its productive
capabilities. Following \cite{im}, a metabolite $\mu$ is said to be
{\it producible} from a given set of nutrients (to be specified) if
\begin{equation}\label{pr}
\exists~\boldsymbol{\nu}\geq\mathbf{0}~\text{such that}~~\mathbf{\Xi}\boldsymbol{\nu}\geq 0~\&~\left[\mathbf{\Xi}\boldsymbol{\nu}\right]_\mu>0
\end{equation}
(where
$\left[\mathbf{\Xi}\boldsymbol{\nu}\right]_\mu=\sum_{i=1}^N\nu_i\xi_i^\mu$)
that is, if at least one flux vector exists allowing for a net
production of $\mu$ irrespective of whether other metabolites are
being also produced (nutrient usage may evidently never exceed its
supply). Producible metabolites have the property that their
concentrations can increase in a stationary flux state with the sole
consumption of the nutrients, so that the cell is allowed to employ
them for purposes other than metabolic (e.g. to form proteins,
membranes, etc.). One would therefore expect that survival in a given
environment is given by the ability to produce the necessary
metabolites be they biomass components or else.

It is easy to understand that the possibility to actually produce a
metabolite may be limited by the emergence of conservation laws from
the stoichiometry. To clarify, note that $M$-dimensional vectors
$\mathbf{z}\geq \mathbf{0}$ satisfying
$\mathbf{z}^T\mathbf{\Xi}=\mathbf{0}$ (or, positive semidefinite
vectors from the left null-space of the stoichiometric matrix) define
linear combinations of metabolites such that the corresponding
weighted sums of their concentrations are constant over time
\cite{fam}. Such conserved moieties are abundant in real metabolic
networks \cite{moy}. Clearly, an accumulation of metabolites belonging
to one such pool, and hence their producibility, is ruled out by
simple stoichiometric reasons. The duality of producibility and
conservation can be exploited to identify (by linear programming) the
growth media allowing for the production of specified sets of
metabolites such as the biomass or slight modifications of it
\cite{belta}\footnote{An interesting extension of producibility is
  {\it sustainability}, loosely defined (see \cite{oe}) as the
  property of being producible using producible metabolites as
  substrates besides nutrients.}.

In order to evaluate the robustness of the cellular production profile
(if any) emerging in given nutrient conditions one should study the
set
\begin{equation}\label{set}
\mathcal{V}=\{ \boldsymbol{\nu}~\text{such that}~\mathbf{\Xi}\boldsymbol{\nu}\geq 0 \}
\end{equation}
(we shall henceforth assume that $\sum_i\nu_i=N$). Statistical
sampling in this case turns out to be feasible, a possible route being
found in the work of J. Von Neumann. In order to present a rather
simplified version of his model \cite{jvn}, let us re-define the
stoichiometric matrix by separating the matrix $\mathbf{B}$ of input
coefficients from the matrix $\mathbf{A}$ of output coefficients, so
that $\mathbf{A}-\mathbf{B}=\mathbf{\Xi}$. It is simple to see
\cite{dm} that, given a constant $\rho>0$, a flux vector
$\boldsymbol{\nu}$ such that $\mathbf{A}\boldsymbol{\nu}\geq \rho
\mathbf{B}\boldsymbol{\nu}$ describes a network state in which every
species is being produced at a rate at least equal to $\rho$. In a
generic input-output system specified by matrices $\mathbf{A}$ and
$\mathbf{B}$, one can therefore ask what is the maximum value of
$\rho$ for which flux vectors satisfying
\begin{equation}\label{vn}
(\mathbf{A}-\rho\mathbf{B})\boldsymbol{\nu}\geq \mathbf{0}
\end{equation}
exist. This value (denoted here as $\rho^\star$) represents the
optimal productive performance allowed by the ways in which the
available processes combine the metabolites. Depending on whether
$\rho^\star$ is larger or smaller than 1, the system may have optimal
states that are expanding or contracting. Correspondingly, the flux
vector(s) satisfying the above set of constraints for
$\rho=\rho^\star$ are the optimal flux states of the system. When
$N,M\to\infty$ and $\mathbf{A}$ and $\mathbf{B}$ are taken to be
semi-positive but otherwise unstructured random matrices, one can
dissect the phase structure of the problem in detail by statistical
mechanics tools, specifically using the replica trick (in a fully
connected network where each reaction consumes and produces a finite
fraction of the possible metabolites \cite{dm}) or the cavity method
(in a finitely connected network where reactions use a finite number
of substrates to generate a finite number of products \cite{ran}). The
general lesson is that a critical value $n_c$ of the structural
parameter $n=N/M$ separates contracting from expanding regimes, with
enhanced dilution increasing (resp. decreasing) $\rho^\star$ in the
expanding (resp. contracting) phase. Moreover, in such cases a single
flux vector satisfies conditions (\ref{vn}) when $\rho=\rho^\star$. In
the biologically more sensible case in which $\mathbf{A}$ and
$\mathbf{B}$ are real stoichiometric matrices, the situation is
radically different. The mass balance conditions imposed by
stoichiometry or, equivalently, the existence of conserved metabolic
pools, imply $\rho_\star=1$ (and force a finite volume of solution in
a random model with realistically structured input and output matrices
\cite{epl}), so that the optimal solution space of (\ref{vn}) indeed
coincides with (\ref{set}). Exploiting the presence of $\rho$ it is
however possible to define a straightforward iterative algorithm that
samples $\mathcal{V}$ uniformly \cite{ran}. This allows for a complete
and extensive characterization of producible metabolites and of the
corresponding flux states in real cellular networks in a given
environment at a modest computational cost even for genome-scale
systems.

\section{Application to E.coli: production profiles and their fluctuations}

The volume (\ref{set}) generated by the stoichiometric matrix of the
bacterium {\it E.coli} has been studied in \cite{pnas}, revealing that
\begin{enumerate}
\item[(a)] the predicted ranges of variability of the fluxes in
  $\mathcal{V}$ in specified extracellular conditions agree well with
  the (limited) experimental data available on the reaction rates
  inferred from $^{13}C$-based experiments in a similar nutrient
  profile \cite{emm,sau};
\item[(b)] dynamically stiff variables, i.e. reactions with smaller
  allowed ranges, tend to correspond, via the associated enzyme, to
  {\it E.coli}'s phenomenologically essential genes, i.e. genes that
  are both necessary for the organism's survival and highly conserved
  across different bacterial species \cite{gerdes}.
\end{enumerate}
These results suggest that at least in some conditions metabolic
networks may operate close to their optimal productive capacity and
that the ``shape'' of $\mathcal{V}$ may contain useful information
relating the metabolic phenotype to the underlying genotype. The
natural question to ask now is whether a robust metabolite production
profile emerges and how it correlates with the physiologically defined
biomass (in suitable growth media). We therefore take a closer look at
the set of produced metabolites for {\it E.coli} in a minimal growth
medium, an environment with a tunable supply of a limited set of
nutrients formed by inorganic phosphate (PI), O$_2$, SO$_4$, CO$_2$,
K, NH$_3$ and a carbon source (glucose in this case) \cite{almaas}
that resembles closely the M9 medium widely used in the experimental
literature \cite{emm}. This network consists of 1057 reactions
involving 631 metabolites altogether (after pruning the trivial
producibility constraints, see \cite{pnas} for details).

For simplicity, we henceforth set
$\boldsymbol{\lambda}=\mathbf{\Xi}\boldsymbol{\nu}$ and label the flux
vectors from $\mathcal{V}$ as $\boldsymbol{\nu}_\alpha$,
$\boldsymbol{\nu}_\beta$, etc. assuming to have sampled $S$ such
configurations. We shall furthermore write
$\boldsymbol{\lambda}_\alpha=\mathbf{\Xi}\boldsymbol{\nu}_\alpha$.

\begin{figure}[!]
\begin{center}
\includegraphics[width=9cm]{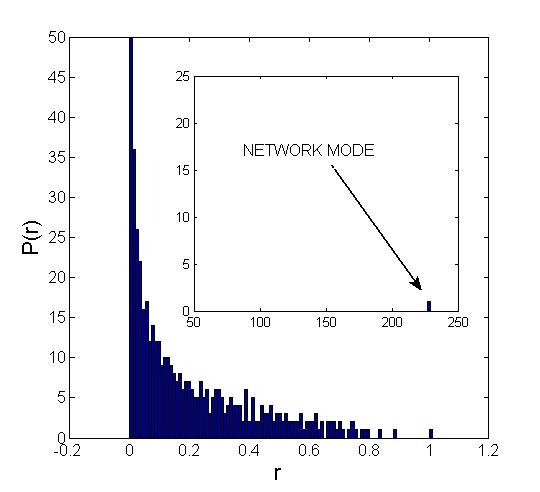}
\includegraphics[width=13cm]{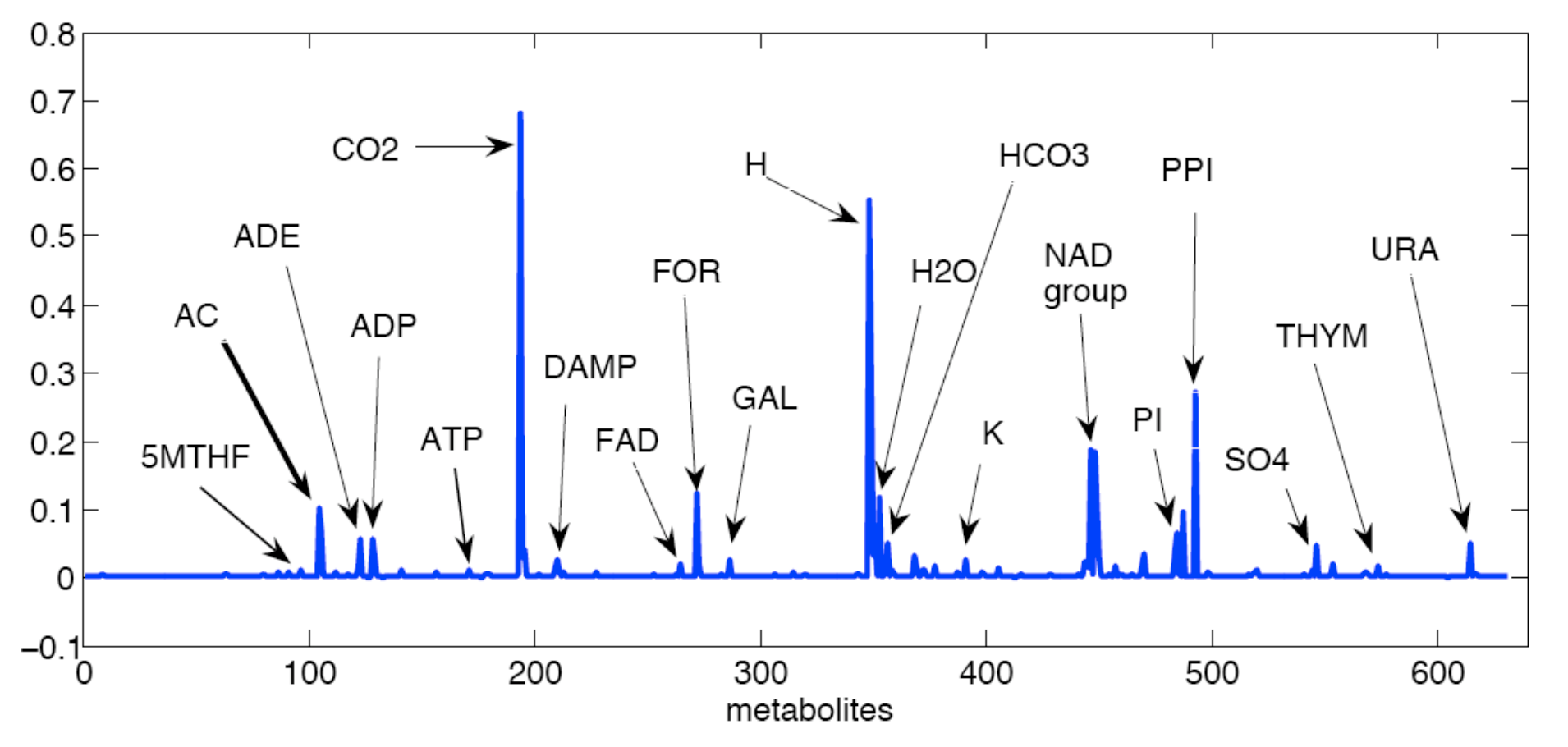}
\caption{\label{bla}(Top) Histogram of the eigenvalues of the
  empirical correlation matrix $\mathbf{C}$, Eq. (\ref{C}), for {\it
    E.coli} in a minimal growth medium (see text). Inset: detail of
  the ``network mode''. (Bottom) Components of the eigenvector of
  $\mathbf{C}$ corresponding to the network mode, with some of the
  producible metabolites explicitly indicated (see \cite{biom2} for a
  list of metabolite abbreviations).}
\end{center}
\end{figure}

The simplest way to check whether $\mathcal{V}$ defines a set of
consistently produced metabolites is by studying the empirical
correlation matrix $\mathbf{C}$ given by
\begin{equation}\label{C}
C^{\mu\mu'}=\frac{1}{S}\sum_{\alpha=1}^S \lambda^\mu_\alpha \lambda^{\mu'}_\alpha
\end{equation}
In the spirit of principal component analysis, $\mathbf{C}$'s
eigenvalue spectrum contains much information on the collective
properties of the system. In particular, it is possible to express the
underlying ``signals'' $\{\boldsymbol{\lambda}_\alpha\}$ as weighted
sums of the eigenvectors of $\mathbf{C}$. The presence of a large,
isolated eigenvalue $r_{max}$ indicates that $\lambda^\mu_\alpha\simeq
\sqrt{r_{max}} ~V_{max}^\mu \eta_\alpha$, where $\mathbf{V}_{max}$ is
the eigenvector corresponding to $r_{max}$ and $\eta_\alpha$ is a
unit-variance random variable. Hence $r_{max}$ and its corresponding
eigenvector may offer an effective, zero-order description of
$\mathcal{V}$ in terms of the emergent production profile. The
eigenvalue distribution for {\it E.coli} is shown in
Fig. \ref{bla}\footnote{To remedy the uncertainty about relative
  fluctuations when $\lambda^\mu\ll 1$, one can set a threshold
  $\epsilon$ below which $\lambda^\mu$ is ``effectively'' zero. For
  the present case, values of $\epsilon\leq 0.01$ have been found to
  provide the same qualitative picture.}.
One clearly sees a giant eigenvalue separated by a roughly continuous
spectrum, well described by the Marcenko-Pastur law with a point mass
at $r=0$ \cite{bouc}. In metabolic terms, $r_{max}$ represents a
``network mode'', a production profile that is common (to a first
approximation) to all $\boldsymbol{\nu}\in\mathcal{V}$. The components
of the eigenvector corresponding to the network mode (Fig. \ref{bla},
bottom) point explicitly to some of the metabolites that are expected
to form the core of the metabolic production. These turn out to
include\footnote{See \cite{biom2} for a list of metabolite
  abbreviations.}
\begin{itemize}
\item biomass components (e.g. 5mthf, atp, fad, h2o, nad, nadh, nadp, nadph)
\item biomass products (adp, h, pi, ppi)
\item end points of metabolic pathways (waste, e.g. ac, co2)
\end{itemize}

\begin{figure}
\begin{center}
\includegraphics[width=13cm]{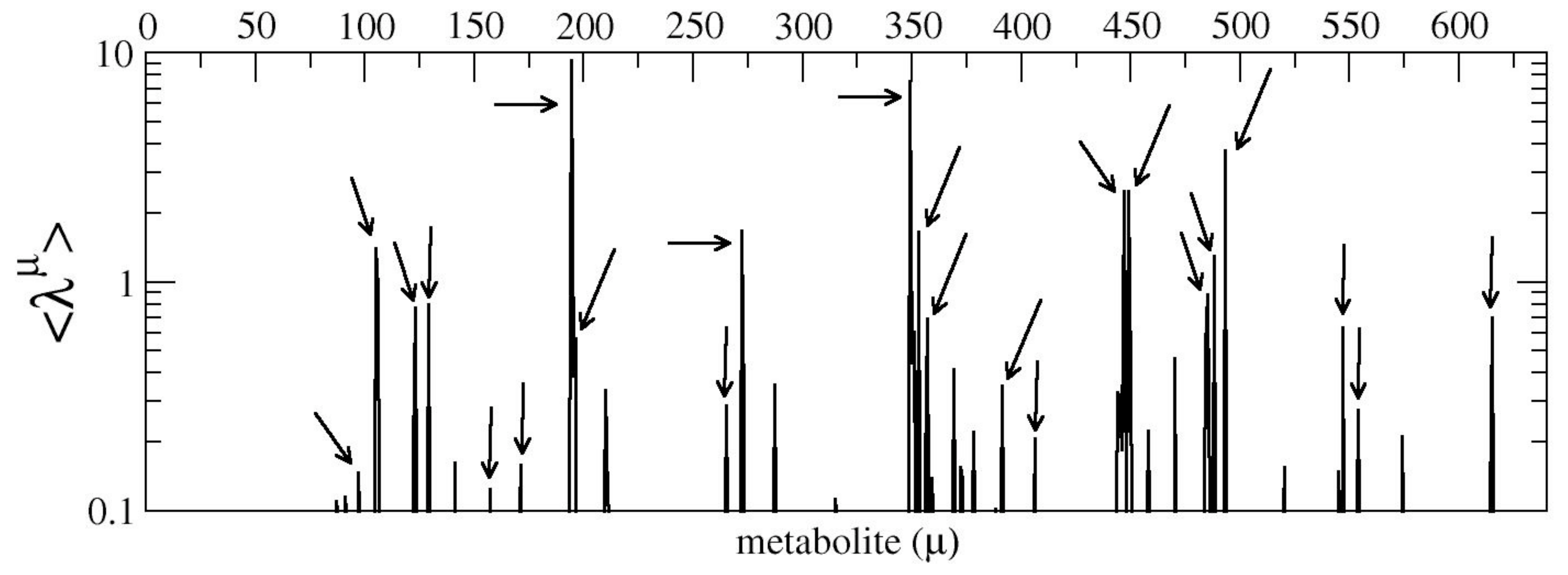}
\caption{\label{uno}Value of $\lambda^\mu$ averaged over 500 flux
  states $\boldsymbol{\nu}\in\mathcal{V}$ versus $\mu$ (metabolite,
  horizontal axis) for {\it E.coli} (631 metabolites) in a minimal
  growth medium (see text). Arrows mark, from left to right, the
  following metabolites (see \cite{biom2} for a list of metabolite
  abbreviations): 5mthf, ac, ade, adp, amp, atp, co2, coa, fad, for,
  h, h2o, hco3, k, leu-L, the nad group (nad, nadh, nadp, nadph), nh4,
  pi, ppa, ppi, so4, succ and ura.}
\end{center}
\end{figure}

Such a profile is largely recovered by studying the average of
$\lambda^\mu$ over 500 flux configurations
$\boldsymbol{\nu}\in\mathcal{V}$ for each $\mu$, see Fig. \ref{uno}.
Focusing for clarity on the ones with the largest average
producibility we find multiple biomass components (5mthf, amp, atp,
coa, fad, h2o, leu-L, nad, nadh, nadp, nadph, ppi) and biomass
products (adp, h, pi and ppi), supporting a growth scenario; typical
waste products of metabolic activity (ac and co2, end point of
oxidative phosphorylation); and metabolites with clear metabolic roles
like formate (central for h production), hco3 and k (helping to
maintain a stable intracellular condition), succ (succinate, a key
node of the Krebs cycle). In addition to these, a host of producible
species appear, though with smaller $\lambda^\mu$, including all amino
acids (with the exception of leucine).

\begin{figure}
\begin{center}
\includegraphics[width=12cm]{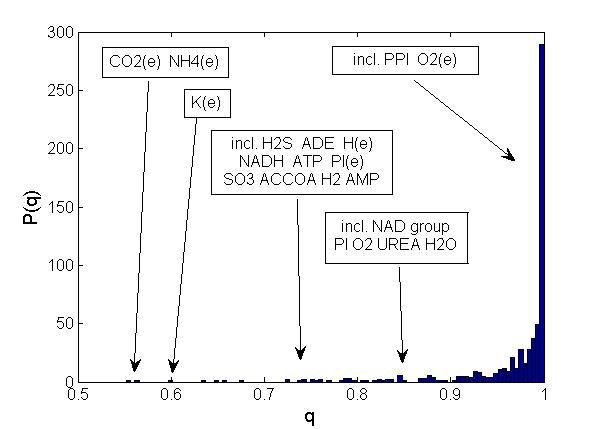}
\caption{\label{disqmu}Distribution of $q^\mu=\langle
  q^\mu_{\alpha\beta}\rangle$ (average over pairs of distinct
  solutions from $\mathcal{V}$). The index e (extracellular) denotes
  metabolites that also serve as nutrients. Arrow mark groups of
  metabolites (as shown) falling in a certain range of $q$.}
\end{center}
\end{figure}

For most compounds, however, one finds
$[\mathbf{\Xi}\boldsymbol{\nu}]_\mu=0$ for each
$\boldsymbol{\nu}\in\mathcal{V}$. As said above,
$[\mathbf{\Xi}\boldsymbol{\nu}]_\mu=0$ if $\mu$ belongs to a conserved
metabolic pool. Interestingly, such moieties do not seem to exhaust
the list of unproducible metabolites, as metabolite producibility
turns out to be a significantly fluctuating property. To evaluate how
the production profile of each metabolite varies across solutions in
$\mathcal{V}$ one can calculate, for each $\mu$ and for each pair of
(distinct) flux vectors
$\boldsymbol{\nu}_\alpha,\boldsymbol{\nu}_\beta\in\mathcal{V}$, the
overlap
\begin{equation}
q^\mu_{\alpha\beta}=\frac{2 \lambda^\mu_\alpha \lambda^\mu_\beta}{(\lambda^\mu_\alpha)^2+(\lambda^\mu_\beta)^2}
\end{equation}
When the above quantity is averaged over pairs $\alpha\neq\beta$
(properly accounting for the case where at least one
$\lambda_\alpha^\mu\ll 1$), one obtains an index (we shall denote it
by $q^\mu$) that is closer to $1$ the smaller are the fluctuations in
$\lambda^\mu$ for metabolite $\mu$. The distribution of $q^\mu$ is
shown in Fig. \ref{disqmu}.
One sees that the production rate of many key metabolites including
co2, atp, amp, the nad group, etc. may oscillate considerably from
solution to solution in $\mathcal{V}$. This holds for intracellular
species (e.g. ATP or biomass components) as well as for extracellular
ones, so that even the amount of nutrients necessary for the cell to
maintain its productive capability in an environment where a specific
group of metabolites is present may have a large allowed range.

\begin{figure}
\begin{center}
\includegraphics[width=15cm]{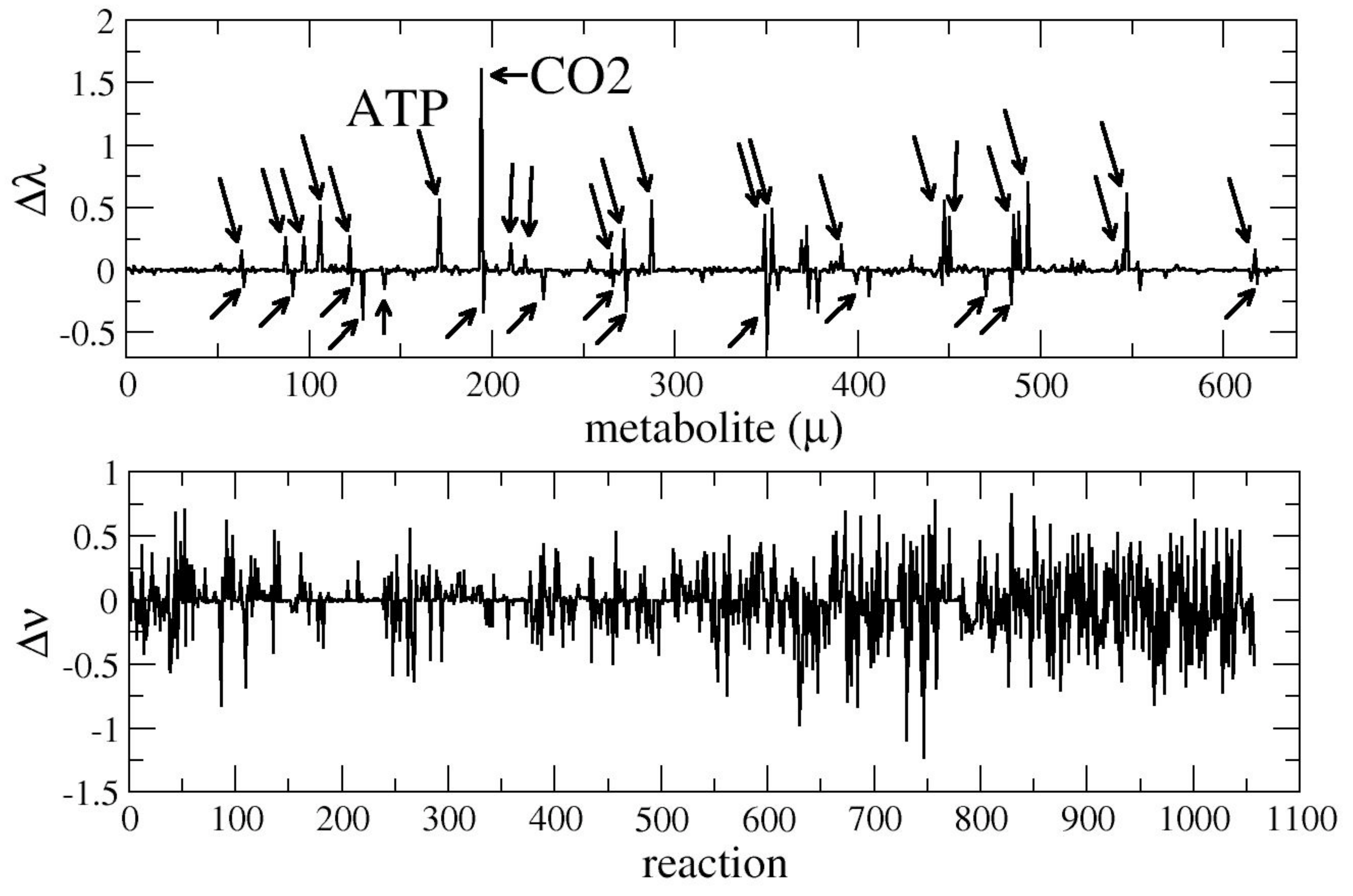}
\caption{\label{due}(Top) Component by component difference in
  $\boldsymbol{\lambda}$ and (Bottom) in $\boldsymbol{\nu}$ between
  the solution with the largest and smaller production rate of
  ATP. The number of metabolites is 631, that of reactions
  1057. Arrows in the top figure mark, from left to right, the
  following metabolites: in the positive half, 3mob, 5aizc, 5mthf, ac,
  ade, atp, co2, damp, dcyt, fad, for, gal, h, h2o, k(e), nad group,
  nh4, pi(e), ppi, so3, so4(e) and urea; in the negative half 3mop,
  5caiz, ade(e), adp, akg, co2(e), dhf, fadh2, for(e), h(e), lac,
  pant-R, pi and urea(e).}
\end{center}
\end{figure}

It is instructive to inspect two particular solutions out of those
sampled from $\mathcal{V}$, namely those with the largest and smaller
production of ATP, respectively. Denoting these respectively by
$\boldsymbol{\nu}_{max}$ and $\boldsymbol{\nu}_{min}$, in
Fig. \ref{due} we show, component by component, the quantities
$\Delta\boldsymbol{\lambda}=\boldsymbol{\lambda}_{max}-\boldsymbol{\lambda}_{min}$
and
$\Delta\boldsymbol{\nu}=\boldsymbol{\nu}_{max}-\boldsymbol{\nu}_{min}$.
Analyzing the production profiles it turns out that the state with
large ATP production is closely associated to the network mode,
i.e. it comes with a significant production of biomass components and
of metabolic waste like ac and co2. On the contrary states with
reduced ATP production are associated to considerably decreased co2
output and increased lactate output. This is consistent with an
increase of the flux through glycolisis, a central, anaerobic carbon
pathway with limited ATP productive efficiency that leads to the
generation of lactate. Large ATP output is instead achievable by
strengthening aerobic pathways that employ oxidative phosphorylation
with the concomitant production of CO$_2$. While a more careful
analysis is needed to map the exact locations of the flux
rearrangements, the overall difference in the organization of flux
between the two states is indeed significant. From a biological
perspective, these states are likely to lead to remarkably different
growth properties, suggesting that a sharper selection of production
profiles, and thus a better identification of metabolic objectives,
requires constraints that are still not included in the available
theories.

\section{Discussion}
Metabolic networks control, directly or indirectly, many of the most
basic tasks cells must accomplish, from the synthesis of amino acids,
to the maintenance of osmotic balance with the exterior, to the
response to environmental shifts. A possible key to improve our
understanding of their organization lies in our opinion in the
identification of the physical, stoichiometric, thermodynamic or
regulatory factors that intrinsically limit their productive potential
and determine the overall outcome of metabolic
activity. Constraint-based approaches provide a simple mathematical
frameworks where many of the emerging properties of metabolism can be
analyzed quantitatively. We have shown here that while the constraints
that are normally employed are sufficient to describe the main
physiological aspects of a cell's growth performance in a certain
medium, they may be still insufficient to infer precise metabolic
objective functions by which one could capture, e.g., objective shifts
under varying nutrient conditions. Otherwise it would be important to
understand whether the observed fluctuations have biological
counterparts. In addition to the problems presented here, the methods
of statistical mechanics developed for the analysis of
constraint-satisfaction problems may prove crucial to address many of
the important questions arising in this field, both for the structural
and the dynamical level \cite{rbc}.

\ack 
This work was supported by the IIT (Italian Institute of Technology)
through the Seed Project DREAM.

\end{document}